\documentstyle[aps,epsfig,twocolumn]{revtex}

\begin{document} 
\draft 
\flushbottom 
\twocolumn[\hsize\textwidth\columnwidth\hsize\csname 
@twocolumnfalse\endcsname 
 
\title{Effect of impurities on quasi-2D quantum antiferromagnet} 
\author{A. L. Chernyshev$^{1,\dag}$, 
Y. C. Chen$^{2}$, and A. H. Castro Neto$^{3,*}$} 
\address{
$^{1}$Solid State Division, Oak Ridge National Laboratory, Oak Ridge, 
Tennessee 37831\\ 
$^{2}$Department of Physics, University of California, Riverside, 
California 92521\\ 
$^{3}$Department of Physics, Boston University, Boston, MA 02215} 
\date{\today} 
\maketitle 
 
\widetext\leftskip=1.5cm\rightskip=1.5cm\nointerlineskip\small 
\begin{abstract} 
\hspace*{2mm} 
We have studied the static and dynamic properties of
quasi-two-dimensional quantum antiferromagnets (AF) diluted with
spinless impurities using spin-wave theory and $T$-matrix
approximation. We show that 
the spectrum of a 2D AF at long wavelengths is overdamped
at arbitrary concentration of spinless impurities. The
scattering leads to a new length scale $\ell/a \sim
e^{\pi/4x}$, $x$ being impurity concentration
and $a$ the lattice spacing, beyond which the influence of
impurities on the spectrum is dominant.  
Although the dynamical properties are
significantly modified we show that 2D is not the lower critical
dimension for this problem. Thus, in low-dimensional systems with
disorder the connection between static and dynamic quantities is not
straightforward. Our results are in quantitative agreement with the
recent Monte Carlo simulations and experimental data for $S=1/2$,
$S=1$, and $S=5/2$. We have also proposed experiments
which can further test the results of our theory.  
\end{abstract} 
\pacs{PACS numbers: 75.10.-b, 75.10.Jm, 75.10.Nr, 75.40.Gb} 
] 
\narrowtext 
 
\section{Introduction} 
 
The discovery of superconducting cuprates has attracted
much attention to the properties of diluted 2D,  QHAF
\cite{NN,KK_imp,Sandvik,kampf,Cheong,Hucker,Carretta,Clarke,Kato,anders}, 
the subject intensively studied
some 30 years ago in the context of magnetism in diluted magnetic alloys 
\cite{Harris,stinchcombe}. 
Traditional view of the effect of local  
disorder on the spectrum of an ordered 3D antiferromagnet is that at 
long wavelengths the {\it form} of the spectrum is not modified.  
The only effects are the reduction of hydrodynamic parameters 
and a weak damping.  
 
In this work we study the problem of impurities in 2D QHAF within the linear 
spin-wave theory using the $T$-matrix approach. 
 The spin-wave Green's function is evaluated by 
summing all multiple-scattering diagrams that involve single 
impurity. 
We find that the scattering 
leads to a new length scale  
$\ell/a \sim e^{\pi/4x}$  
beyond which the influence of impurities on the spectrum is dominant.  
We associate this length scale with the localization length of spin 
excitations. 
 We show that the dynamical structure factor ${\cal S}({\bf k},\omega)$ for 
$a^{-1}\gg k\gg\ell^{-1}$ consists of three
parts (we use units $\hbar=k_B=1$): 
({\it i}) a broadened quasiparticle  
peak  with a width  
given by $\Gamma_k \simeq x \, c_0 k$, $c_0= 2 \sqrt{2} S J a$;  
({\it ii}) a non-Lorentian localization 
peak at  $\omega =\omega_0 \sim c_0\ell^{-1}$, ({\it iii}) a flat
background of states between $\omega=c_0 k$ and $\omega=\omega_0$.  
Thus for every ${\bf k}$-state some weight is spread  from the high energies
 to the low energies. 
For $k \alt \ell^{-1}$ the quasiparticle and localization peaks 
 merge into a broad incoherent peak that
disperses in momentum space.
We calculate the static magnetic properties and find a quantitative 
agreement  
with both MC simulations and experimental data. We show that at $T=0$ the 
staggered magnetization
is given by $M(x,0) \approx S-\Delta-B x$, 
the factor $\Delta\approx 0.2$ is the contribution of the 
zero-point fluctuations of the spins and
$B\simeq 0.21$. 
We find that $T_N(x)/T_N(0) \simeq 1 - A_s\, x $  where 
$A_s=\pi-2/\pi +B/(S-\Delta)$.  
This result gives $A_{1/2}\simeq 3.2$ and 
$A_{5/2}\simeq 2.6$ which agree very well with experiments
up to $x\simeq 0.25$.  
 
The systems discussed in this paper are modeled by the site-diluted 
 quantum Heisenberg antiferromagnet:  
\begin{eqnarray} 
H=\sum_{\langle ij\rangle} J_{ij} \, p_{i} \, p_{j} \, {\bf S}_{i}  
\cdot {\bf S}_{j} \, ,   
\label{H1} 
\end{eqnarray} 
where $p_i=1$ ($0$) if $i's$ site is occupied (unoccupied) by the spin 
$S$. We focus on the problem of tetragonal or square lattices with in-plane, 
$J$, and out-of-plane, $J_{\perp}$, nearest-neighbor exchange 
constants, $\langle ij\rangle$ denotes summation over bonds.   
In the systems of interest $J \gg J_{\perp}$. 
We begin with the Hamiltonian (\ref{H1}) in the linear spin-wave
approximation which is split into the pure host and impurity part 
and, 
after the Bogolyubov transformation, is given by (in the units of $4SJ$)
\begin{eqnarray} 
\label{H_BT_0} 
&&{\cal H}_0=\sum_{\bf k} \omega_{\bf k} \left(\alpha^{\dag}_{\bf k} 
\alpha_{\bf k}+\beta^{\dag}_{\bf k} 
\beta_{\bf k}\right)\ ,\\ 
\label{H_BT} 
&&{\cal H}_{imp}=-\sum_{l,{\bf k},{\bf k}^\prime} 
e^{i({\bf k}-{\bf k}^\prime){\bf R}_l} {\hat{\cal A}}_{\bf k}^{\dag} 
{\hat{\cal V}}^l_{{\bf k},{\bf k}^\prime} {\hat{\cal A}}_{{\bf k}^\prime} \ , 
\end{eqnarray} 
with two-component vectors ${\hat{\cal A}}_{\bf k}^{\dag} 
= \left[ \ 
\alpha_{\bf k}^{\dag}, \
\beta_{-{\bf k}} 
\ \right]$
and $2\times 2$ scattering potential matrices 
${\hat{\cal V}}_{{\bf k},{\bf k}^\prime}$ (for details see
\cite{Chernyshev}), $l$ runs over the impurity sites.  
We are interested in the Green's function of the Hamiltonian 
(\ref{H_BT_0}) modified by random impurity potentials 
(\ref{H_BT}). The Green's function is a $2\times 2$ matrix defined in 
a standard way and the 
$T$-matrix equation for the Hamiltonian (\ref{H_BT}) is given by: 
$\hat{T}^{l,\mu}_{{\bf k},{\bf k}^\prime}(\omega)  
=-{\hat{\cal V}}^{l,\mu}_{{\bf k},{\bf k}^\prime}  
-\sum_{\bf q}{\hat{\cal V}}^{l,\mu}_{{\bf k},{\bf q}} \hat{G}^0_{\bf q} 
(\omega) \hat{T}^{l,\mu}_{{\bf q},{\bf k}^\prime}(\omega)$, 
with $l=A(B)$, partial waves are restricted to in-plane  
$\mu = s,p_\sigma,d$ harmonics  
and $\hat{G}^0_{\bf q}(\omega)$ is the 
bare Green's function.
The $T$-matrix equations can be solved and the subsequent 
averaging over random distribution of impurities  
transforms $T$-matrix into the spin-wave 
self-energies $\hat{\Sigma}_{\bf k}(\omega)= 
x \delta_{{\bf k}-{\bf k}^\prime}
\sum_\mu \left[\hat{T}^{A,\mu}_{{\bf k},{\bf 
k}^\prime}(\omega)+\hat{T}^{B,\mu}_{{\bf k},{\bf k}^\prime}(\omega) 
\right]$.
Summation of the Dyson-Belyaev  
diagrammatic series for the Green's functions 
 gives $\hat{G}=\hat{G}^0+\hat{G}^0\hat{\Sigma}\hat{G}$ and
then the transverse component of
the neutron scattering dynamical structure  
factor is proportional to the linear combination of the
spectral functions 
\begin{eqnarray} 
\label{S_perp} 
{\cal S}^\perp({\bf k},\omega)=f({\bf k}) g(\omega) 
\left[ A^{11}_{\bf k}(\omega)+  \ 
A^{22}_{\bf k}(\omega)+ 2 A^{12}_{\bf k}(\omega)\right]\ ,\nonumber 
\end{eqnarray} 
where $A^{ij}_{\bf k}(\omega)=-\frac{1}{\pi} 
\mbox{Im}{\hat{G}}^{ij}_{\bf k}(\omega)$,
kinematic form-factor $f({\bf k})
=\pi\, S\, (1-\gamma_{\bf k})/\omega_{\bf k}$, 
$g(\omega)=[1+n_B(\omega)]$, and
$n_B(\omega)=[e^{\omega/T}-1]^{-1}$ is the
Bose distribution function.  
 
The static properties of the system 
are calculated from the spin-wave expression 
of the averaged on-site magnetic moment: 
\begin{eqnarray} 
\label{Sz} 
|\langle S^z_i\rangle|=S+\frac{1}{2}-\sum_{\bf k}\frac{1}{\omega_{\bf k}}
\left[ \frac{1}{2}+\langle
\alpha_{\bf k}^{\dag}\alpha_{\bf k}\rangle -\gamma_{\bf k} 
\langle \alpha_{\bf k}^{\dag}\beta_{\bf k}^{\dag}\rangle\right] 
\nonumber \ , 
\end{eqnarray} 
where bosonic averages can be expressed through the spectral 
functions. 

\section{Results} 
\label{dynamic} 

\noindent
\begin{figure}
\unitlength 1cm
\begin{center}
\epsfig{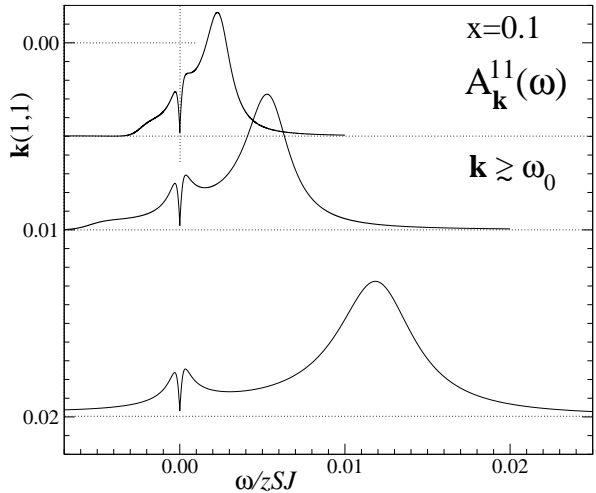}
\end{center}
\caption{The spectral function $A^{11}_{\bf k}(\omega)$ 
for the wave-vectors ${\bf k}=0.005, 0.01,$ and $0.02$
along the $(1,1)$ direction, ${\bf k}=0.005$ is of the order of 
$\omega_0$.  $A^{11}_{\bf k}(\omega)$ 
for each ${\bf k}$ is normalized to fit the picture.} 
\label{fig_1}
\end{figure}
\noindent
A detailed analysis of the spectral function gives the following picture.
At the wave-vectors much larger than $\omega_0$ ($\omega_{\bf
k}\gg\omega_0$), that is at the wavelengths shorter than a characteristic
length $\ell\sim e^{-\pi/4x}$, the spectral function has three
distinct regions in $\omega$.
First, is a vicinity of a quasiparticle peak,
$\omega\approx\tilde\omega_{\bf k}$,
where the spectrum has a regular Lorentzian form with 
the pole at $\tilde{\omega}_{\bf k}$ and width $\tilde\gamma_{\bf k}$
given by the perturbative result.
Second, the intermediate range of energies, 
$\omega_0<\omega\ll\tilde\omega_{\bf k}$, where the spectral function
is independent of $\omega$ and corresponds to an almost flat,
shallow ($\sim x$) background of states. 
Third, the vicinity of a ``localization peak'', 
$\omega\approx\omega_0$,
where the spectral function rises sharply from the shallow background 
states $\sim x$ 
to a peak of the height $\sim 1/x$ and then
vanishes in a singular fashion as $\omega$ approaches zero. 
The spectral function $A^{11}_{\bf k}(\omega)$ 
 is shown in
Fig. \ref{fig_1} for a representative value of
impurity concentration $x=0.1$. 
One can clearly see the features we have discussed above: 
the broadened quasiparticle peak,
the localization peak, and the states between them.
As the ${\bf k}$ decreases all the mentioned structures merge.

The average on-site magnetic moment for randomly
diluted AF with the averaging over magnetic sites
$M(x)=\sum_i |S^z_i|/N_m$ can be expressed
through the integral of the spectral functions as:
\begin{eqnarray} 
\label{M} 
&&M(x,T)=S-\Delta-\delta M(x,T)\ ,\\
&&\delta M(x,T)=\sum_{\bf k}
\int_{-\infty}^{\infty}\frac{n_B(\omega)\,\mbox{d}\omega}{\omega_{\bf k}}
\big[ A^{11}_{R,{\bf k}}(\omega)-\gamma_{\bf k}A^{12}_{R,{\bf
k}}(\omega)\big]\ , \nonumber
\end{eqnarray} 
where $\Delta=\sum_{\bf k}v_{\bf k}^2\simeq 0.1966$ is the
zero-point spin deviation, subscript $R$ denotes retarded.

\noindent
\begin{figure}
\unitlength 1cm
\begin{center}
\epsfig{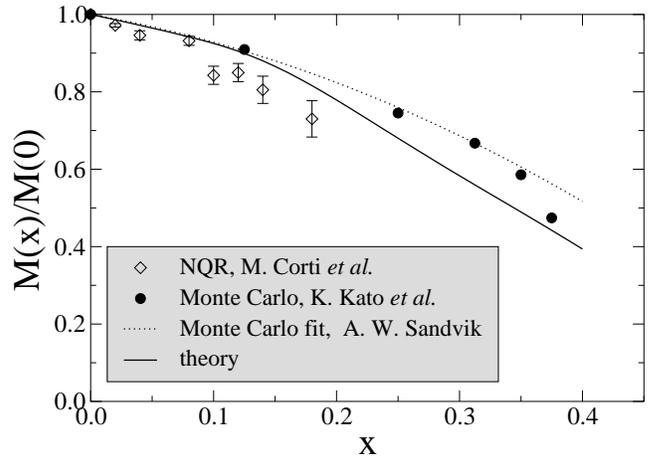}
\end{center}
\caption{Average staggered magnetization v.s. $x$. Our results from
Eq. (\ref{M}) (solid line), Monte Carlo data (circles, Ref.
\protect\cite{Kato}), NQR data (diamonds,
Ref. \protect\cite{Corti}), and the fit of Monte Carlo data from
Ref. \protect\cite{anders} (dotted line) are shown.} 
\label{fig_2}
\end{figure}
\noindent
Numerical integration of the expression in Eq. (\ref{M}) 
 gives the suppression rate of
the staggered magnetization $M(x)\simeq M(0)- Bx$ with $B=0.209(8)$.
For $S=1/2$ it gives the slope of the normalized staggered
magnetization $M(x)/M(0)\simeq
1-Bx/(S-\Delta) \simeq 1-0.691(5)\cdot x$.
The staggered magnetization v.s. $x$ is 
presented in Fig. \ref{fig_2} for $S=1/2$. The recent
Monte Carlo data Refs.\cite{Kato,anders}, 
and NQR data Ref. \cite{Corti} are also shown. One
can see a very good agreement of our results with numerical data up
to high concentrations. 
\noindent
\begin{figure}
\unitlength 1cm
\begin{center}
\epsfig{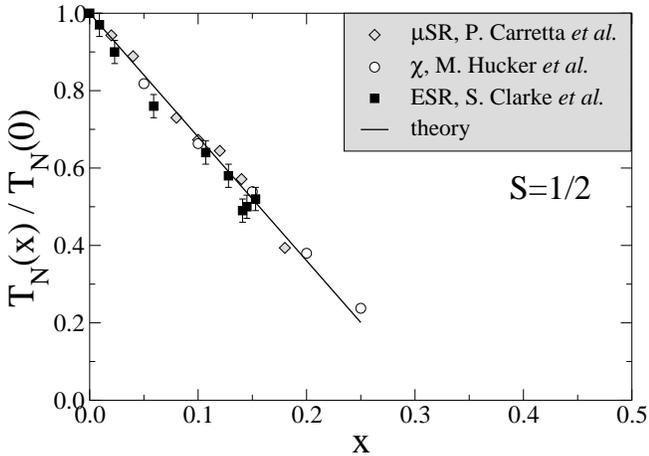}
\end{center}
\caption{$T_N(x)/T_N(0)$ v.s. x for $S=1/2$.
Our results (solid line),
$\mu$SR (diamonds) \protect\cite{Carretta}, magnetic
susceptibility  
(circles) \protect\cite{Hucker}, and ESR (squares) 
\protect\cite{Clarke} data.}
\label{fig_3}
\end{figure}
\noindent
At $T>0$  the staggered magnetic moment consists of the 
quantum, $T=0$, and thermal, $T$-dependent,
parts. For the true 2D system at $x=0$ and $T>0$ thermal fluctuation destroy
the long-range order 
which manifests itself as a log-divergency of the thermal
correction to the magnetization.
The 3D coupling provides a cut-off to this divergency in a quasi-2D
problem which yields the finite value of the thermal correction 
and the finite value of the N\'eel temperature whose mean-field value
can be found from the condition $M(x,T_N)=0=S-\Delta-\delta M(x,T_N)$.
Suppression rate of the N\'eel temperature can be
readily obtained $\frac{T_N(x)}{T_N(0)}\simeq 1-A_s\, x=1-x\bigg(\pi
-\frac{2}{\pi}+\frac{B}{S-\Delta}\bigg)$.
For $S=1/2$ this gives 
$A_{1/2}=3.196(5)$ and for $S=5/2$ it is
$A_{5/2}=2.600(4)$. 
We plot our results for $T_N(x)/T_N(0)$ for the case of
$S=1/2$  in Fig. \ref{fig_3}
together with  experimental
data.  One can see that our 
results agree very well with the experiments up to a rather high
doping level $x\approx 0.25$.

\section{Conclusions}
\label{conclusions} 

We have studied the problem of diluted 2D and quasi-2D
quantum Heisenberg antiferromagnets in a tetragonal lattice
making use of linear spin-wave theory and  $T$-matrix
approach. We have shown 
that the spin-wave spectrum is strongly modified by
disorder which indicates magnon localization on a length scale $\ell$,
exponentially large in $1/x$. This new length-scale
appears explicitly in the dynamic properties such as the dynamical
structure factor ${\cal S}({\bf k},\omega)$
which can be measured directly in
neutron scattering experiments, and the magnon density of states
$N(\omega)$, which 
is directly related to the magnetic specific heat. The measurement
of such quantities will provide a direct test of our theory. Furthermore,
we show that the static properties such as the zero-temperature
staggered magnetization  and N\'eel temperature
 do not show any anomaly associated 
with the spectrum and are finite up to the concentration close to the 
classical percolation threshold. These results are in a quantitative
agreement with the 
NQR, $\mu$SR, ESR, and magnetic susceptibility
measurements in different compounds as well as with the Monte Carlo
data. 
Thus, in low-dimensional systems with disorder
the connection between static and dynamic quantities is not
straightforward. 
Altogether this provides a self-consistent picture of the effects of
disorder in low-dimensional quantum antiferromagnets.

\acknowledgments

This research was supported in part by 
Oak Ridge National Laboratory, 
managed by UT-Battelle, LLC, for the U.S. Department of Energy under
contract DE-AC05-00OR22725, and by a
CULAR research grant under the auspices of the US Department of Energy.
 

\end{document}